\documentclass[12pt,a4paper]{article}
\usepackage{graphicx}
\begin{document}
\noindent\textbf{Toward a complete description of nucleation and growth
in liquid-liquid phase separation}\\
\normalsize
\vspace{0.5cm}
Short title : Nucleation and growth in liquids\\
Jean Colombani$^*$ and Jacques Bert\\
Laboratoire de Physique de la Mati\`ere Condens\'ee et Nanostructures\\
Universit\'e Claude Bernard Lyon 1 and CNRS\\
6, rue Amp\`ere, 69622 Villeurbanne cedex, France\\
$^*$ Corresponding author, Jean.Colombani@lpmcn.univ-lyon1.fr

\section*{Abstract}
The phase separation mechanism of a binary liquid mixture off-critically
quenched in its miscibility gap is nucleation and growth, its homogeneous phase
reaching a metastable equilibrium state.
The successive stages of growth of the nucleated droplets are a diffusion-driven
free growth, an intermediate regime and a coarsening by reduction
of interface (Ostwald ripening or Brownian collisions induced coalescence).
We have made light attenuation experiments to investigate the sedimentation in such
systems.
These results have given us access experimentally to two values predicted
theoretically: the growth exponent of the intermediate regime and the
crossover time between this regime and interface-reduction coarsening.
These data, added to the literature results, have permitted to get a quite complete
view of the growth scenario in very off-critical phase-separating liquids.

\vspace{0.5cm}

\vspace{1cm}

\section{Introduction}
In binary liquid mixtures exhibiting a miscibility gap, the two components experience
an effective repulsive interaction.
At high temperature, this repulsion is overwhelmed by thermal stirring and
the system is monophasic.
Under a composition-dependent temperature (the binodal line,
delimiting the miscibility gap), the entropic mixing is dominated by
the chemical repulsion and the liquid separates in two phases.
This liquid-liquid phase separation gives the opportunity to observe the
pathway of a system evolving from a nonequilibrium or metastable state
toward a stable equilibrium state.
Indeed when quenching the mixture from the one-phase region inside
the two-phases domain, the homogeneous phase abruptly reaches:

\begin{itemize}
\item either a nonequilibrium state, when quenched inside the spinodal line.
The phase separation mechanism is here spinodal decomposition, which is
now quite understood.
\item or a metastable equilibrium state, when quenched between the spinodal
and binodal lines.
The phase-separation process is then nucleation and growth.
If the behaviour of the early and late stages of the growth
are now well established,
the intermediate stage still presents open questions that
we plan to study in this work.
\end{itemize}

Unlike the case of spinodal decomposition where measurements of
the evolution of the growing domain size exist over at least 7 decades of time
\cite{Perrot94},
measurements of the radius $R$ of the growing droplets of the minority phase
after nucleation are available only over narrower windows of time,
as pointed in \cite{Sagui}.
Nevertheless, the collection of the literature results enables to get a view of
the successive processes and corresponding growth laws:

\begin{itemize}
\item After a generally heterogeneous nucleation \cite{Buil00}, the droplets grow with a
$R\sim t^{\frac{1}{2}}$ law \cite{Cumming}.
This evolution corresponds to a "free growth" where the solute migrates through
the bulk phase by molecular diffusion toward the nuclei \cite{Langer}.
\item An intermediate regime of slower growth is subsequently achieved
when the solute-depleted
layers around the nuclei start to overlap and when the interfaces between
the two phases become progressively sharp.
The growth law during this stage and its domain of existence is still a matter
of investigation \cite{Tokuyama}.
\item Afterward, the equilibrium order parameter, i.e., the final volume fraction
of the minority phase, equal to the initial supersaturation $\Phi$, is attained
and the coarsening proceeds through the diffusion of this order parameter via
two possible
mechanisms : either Ostwald ripening (small nuclei evaporate and condensate onto
large ones) \cite{Lifschitz} or coalescence induced by Brownian collisions
\cite{Siggia}.
Both processes lead to a reduction of the total interface and to a
$R\sim t^{\frac{1}{3}}$ law \cite{Wong}.
\item Finally, when the gravitational effects dominate, a macroscopic sharing of the
two phases occurs by sedimentation.
\end{itemize}

In order to evaluate the growth exponent and the range of existence of the
above-mentionned intermediate regime, we have used a sedimentation-based
indirect method which has given us access to the radius evolution
of the droplets over several decades of time.

\section{Experiments}
\begin{figure}[!hbtp]
\begin{center}
\includegraphics[width=8.1cm]{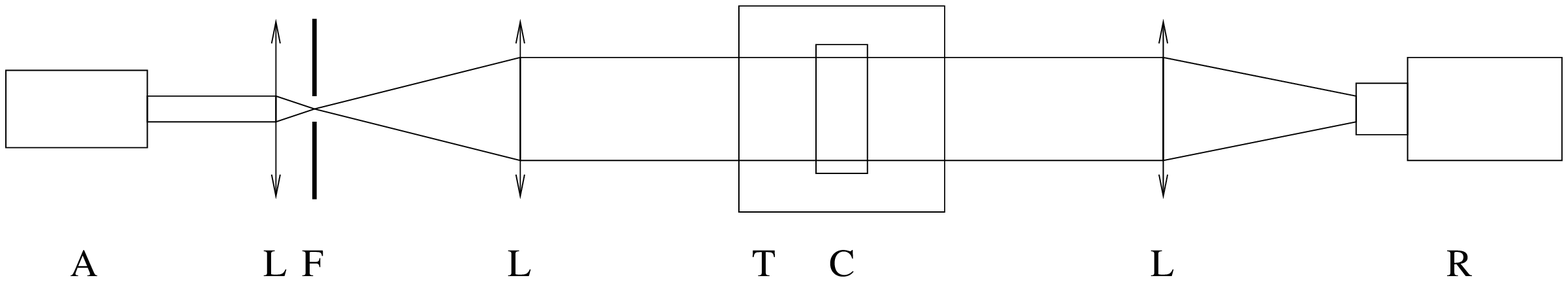}
\vspace{2cm}
\caption[Sketch of the light attenutation experiment: A) Laser (wavelength 532 nm)
L) Converging lens, F) Spatial filter (pinhole), T) Temperature-regulated chamber
(thermal stability 0.01 K), C) Experimental transparent cell (height 3 cm, width
1 cm and optical length 0.1 cm), R) Charge-coupled device camera.]{}
\label{sketch}
\end{center}
\end{figure}

\begin{figure}[!hbtp]
\begin{center}
\includegraphics[width=8.1cm]{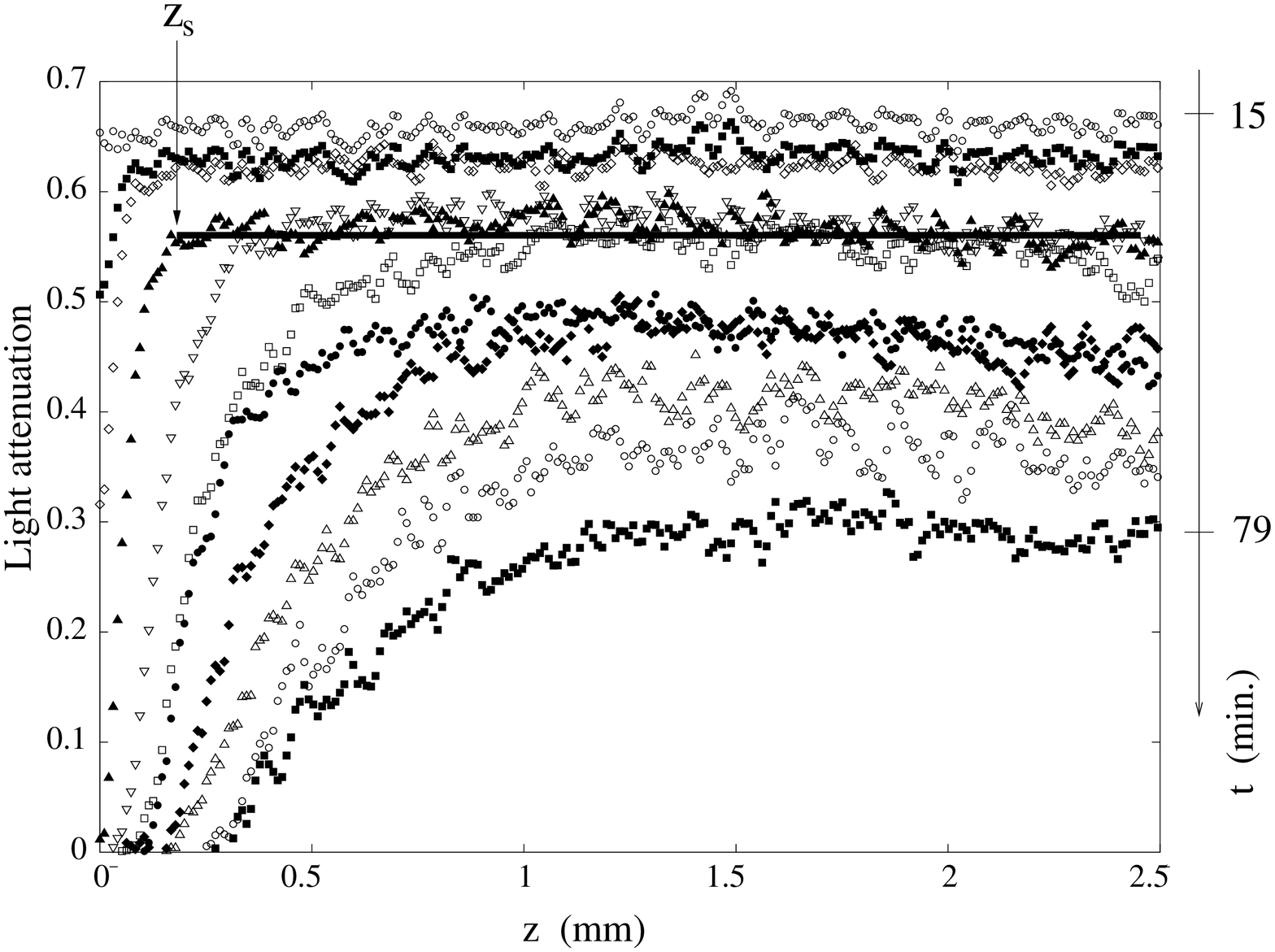}
\vspace{2cm}
\caption[Light attenuation across the cell as a function of position $z$
and time $t$ for an initial supersaturation $\Phi=8.07\%$ and a
quench depth $\delta T=0.35$ K.]{}
\label{absorp}
\end{center}
\end{figure}

\begin{figure}[!hbtp]
\begin{center}
\includegraphics[width=8.1cm]{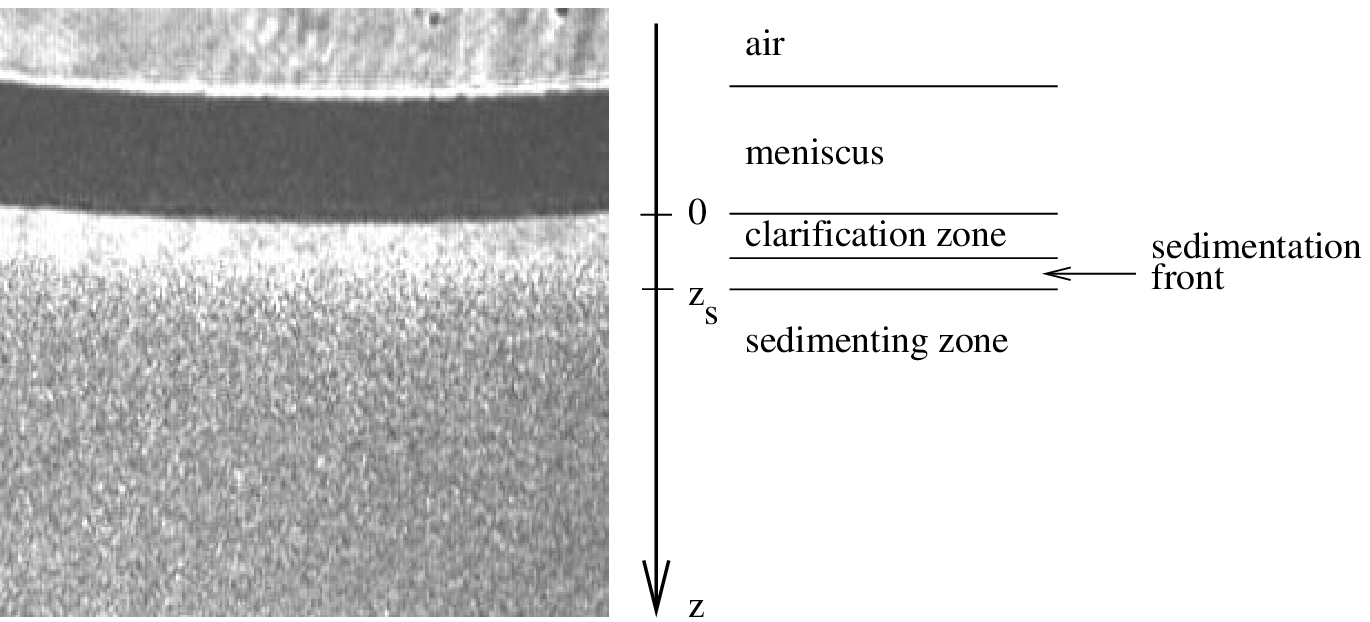}
\vspace{2cm}
\caption[Photograph of the cell during sedimentation 50 min. after the quench
for the experiment of figure \ref{absorp}.]{}
\label{photo}
\end{center}
\end{figure}

Our procedure is the following.
We have quenched off-critical mixtures of water-isobutyric acid (showing a consolute
critical point at $T_c=27.05^{\circ}$C) below the binodal line.
We have then studied the nucleation, growth and sedimentation in the experimental
cell with a light attenuation technique, sketched in figure \ref{sketch}.
Figure \ref{absorp} displays the evolution of the light attenuation curves
(transmitted light intensity over incident light intensity)
along a vertical axis in the cell as a function of time.
Initially the curve is a horizontal line of strong attenuation, which originates in
the scattering of light by the rapidly moving nucleating and growing
droplets of submicrometer size.
When sedimentation sets in, a droplet-free clarification zone appears and expands
at the top of the liquid
due to the average settling motion of all the droplets.
This can be observed in the curves by a steep increase just preceding the plateau:
the number density of droplets starts to decrease at the top, so the light
scattering diminishes at this point.
The bulk droplet-rich liquid and the clarification zone are separated by the
sedimentation front, which broadens with time owing to the droplets
Brownian diffusion (see figure \ref{photo}).

We now focus on the onset of sedimentation at time $t_s$.
The P\'eclet number Pe is the ratio of the buoyant velocity $v$ of a droplet
to its diffusive velocity $D/R$: Pe$=vR/D$ ($D$ droplet diffusivity).
When the former sufficiently exceeds the latter, at $t_s$,
the droplet begins to settle down.
We will tentatively consider that this occurs when the gravity effect overwhelms the
diffusive effect by one order of magnitude, i.e., for Pe$_s=10$.
The position of the first droplets experiencing Pe=Pe$_s$, i.e.,
the first sedimenting droplets, can be traced in figure \ref{absorp}
through the early evolution of $z_s$.
The point $z_s$ is the intercept of the increasing curve with the plateau.
In other words, $z_s$ represents the position of the lower sedimentation front
(cf. figure \ref{photo}) and its motion just at $t_s$ reflects the settling of the
droplets experiencing Pe=Pe$_s$.
The velocity of these droplets is $v_s=(dz_s/dt)_{t=t_s}$.
Their diffusion coefficient is evaluated through a Stokes-Einstein law : $D_s=k_B
T/(5\pi\eta R_s)$ ($\eta$ viscosity of the majority phase).
Therefore the radius of the droplets at the beginning of sedimentation $t_s$,
calculated in introducing the above-mentionned expressions in Pe$_s=v_s R_s/D_s=10$,
is:
\begin{equation}
R_s=\left(\frac{2k_B T}
{\pi\eta\left(dz_s/dt\right)_{t=t_s}}\right)^\frac{1}{2}.
\end{equation}

It should be noticed that the droplets at $t_s$ are at the end of the
"weightless" regime of growth and just enter the sedimentation-influenced growth.
Consequently, their radius at this time derives only from the
sedimentation-free coarsening regimes that they have experienced
until this point (diffusive growth, Ostwald ripening, \ldots) \cite{Colombani}.

To allow comparison with literature results, renormalized values are needed.
The reference size in nucleation and growth mechanisms is the radius
of the critical nucleus
(for which the bulk free energy gain exactly compensates the surface free
energy loss)
$R_c=\alpha/\Phi$ with $\alpha$ a capillary length evaluated as
$\xi^-/3$ ($\xi^-$ is the concentration fluctuations correlation length)
\cite{Langer}.
The reference time is the relaxation time of this critical nucleus
$t_c=R_c^2/(D^-\Phi)$ ($D^-$ solute diffusivity) \cite{Langer}.

\section{Results}
\begin{figure}[!hbtp]
\begin{center}
\includegraphics[width=8.1cm]{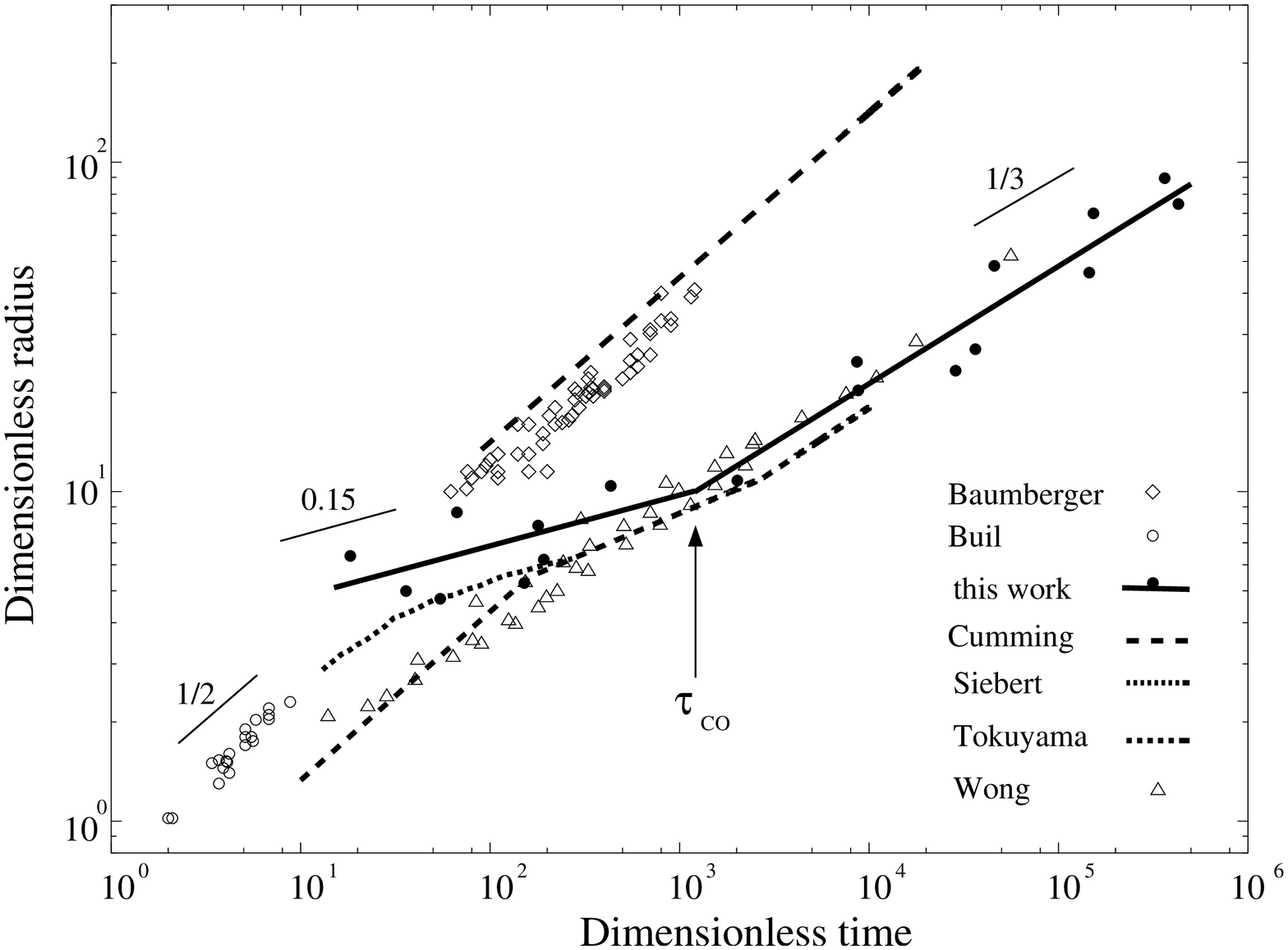}
\vspace{2cm}
\caption[Droplet radius $\rho$ at the onset of sedimentation $\tau$ in reduced units
for experiments with various $\delta T$ and $\Phi<6\%$.
Radius evolutions with time from Baumberger \textit{et al.} \cite{Baumberger},
Buil \textit{et al.} \cite{Buil}, Cumming \textit{et al.} \cite{Cumming},
Siebert and Knobler \cite{Siebert}, Tokuyama and Enomoto \cite{Tokuyama} and
Wong and Knobler \cite{Wong} have also been added.]{}
\label{radius}
\end{center}
\end{figure}

We have then plotted in figure \ref{radius} [$\tau=t_s/t_c$, $\rho=R_s/R_c$] points
for experiments of quench depths $\delta T$ ranging from 0.03 to 0.50 K
and initial supersaturation $\Phi$ ranging from 0.3 to 5.9 \%$_{vol}$.
We note that all these experiments build a universal coarsening behaviour with a
crossover time $\tau_{CO}\simeq 1\times 10^3$ between two growth exponents:
0.15 followed by 0.36 $\simeq$ 1/3.
The latter obviously corresponds to the interface-reduction coarsening,
but the experimental dispersion has not enabled to discriminate between
Brownian coalescence and Ostwald ripening \cite{Colombani}.
It follows that the former characterizes the intermediate regime and that
$\tau_{CO}$ shares the end of the intermediate regime and the beginning
of the conserved order-parameter coarsening.

To check the consistency of this interpretation with known results,
the ($\tau$, $\rho$) curves of the major theoretical and experimental studies
in the literature for very off-critical liquid mixtures, i.e., $\Phi\leq 6$\%,
have been added in figure \ref{radius}.
The discrepancies between these results reveal incoherence so a critical survey
is needed.
Of the pioneering work of Wong and Knobler, only the most off-critical
quenches have been drawn \cite{Wong}.
A crossover dynamics was not expected at this time, the experimental curves
are bent and growth exponents are difficult to evaluate.
The quench numbered $L$ of another study of Wong and Knobler \cite{Wong78},
reanalysed by Siebert and Knobler \cite{Siebert},
corresponds to a supersaturation $\Phi=4.2$ \% and has accordingly been drawn.
Their values take place at the end of the diffusive growth and the beginnning
of the intermediate regime.
For this part of their curve, the authors indicate a growth exponent of 0.18, which
is fairly close to ours (0.15).
The outstanding results of Buil \textit{et al.} obtained by an original optical
method for $\Phi=2.6$ \% give access to the very beginning of the phase separation,
including the nucleation and early free growth, and have been added \cite{Buil}.
The results of Cumming \textit{et al.}, also displayed, show the $t^{1/2}$
coarsening law unexpectedly elapsing far beyond $\tau_{CO}$ \cite{Cumming}.
This feature can be a consequence
either of a specific kinetic behaviour of the polymer blends used by these authors
or more likely of the difficulty to compute
renormalized quantities in macromolecular liquids comparable
to those of simpler systems (cf. a discussion of this point in \cite{Lalaude95}).
Baumberger \textit{et al.} also find a free growth regime for 2 \% $<\Phi<$ 4 \%
---with an exponent 0.6--- beyond
$\tau_{CO}$ despite their use of exactly the same system as ours
\cite{Baumberger}.
We have evaluated the P\'eclet number of their droplets for the studied
dimensionless times and we have found $Pe$ ranging between 50 and 1000.
Therefore their experiments should have been performed during the sedimentation
regime, which could explain their high growth exponent.
Finally, Tokuyama and Enomoto have made a systematic model of the
phase separation dynamics after heterogeneous nucleation, to
predict the coarsening laws in the three regimes, in particular
for $\Phi=1$\% \cite{Tokuyama}.
Their intermediate regime exponent (1/4) is slightly larger than our experimental one
(0.15) but their crossover time $\tau_{CO}\simeq 2\times 10^3$ lies
in agreement with ours ($1\times 10^3$).
As they concern only homogeneous nucleation, the $\rho(\tau)$ curves
of the model of Sagui \textit{et al.} were not included \cite{Sagui} .

\section{Conclusion}
By collecting the main literature results and completing them with
an indirect method based on the light attenuation observation of early
sedimentation, we have achieved a coherent universal description of growth
in phase-separating binary liquid mixtures of low supersaturation.
Some minor questions remain, like for instance the exact respective domains
of existence of Ostwald ripening and Brownian coalescence as a function
of time and supersaturation.
Besides, the field of phase separation in complex systems presents still appealing
problems, like for instance the interplay between phase separation, aggregation and
conformational changes in macromolecular solutions, which show wide applications
in life sciences \cite{Manno}.

\newpage

\listoffigures

\newpage

\newpage

\newpage

\newpage

\end{document}